\documentclass[conference]{IEEEtran}

\IEEEoverridecommandlockouts

\usepackage{cite}
\usepackage{amsmath}
\usepackage{cuted}
\usepackage{amsopn}

\usepackage{float}

\usepackage{dblfloatfix}

\usepackage{setspace}

\usepackage{amssymb}

\usepackage{amsfonts}

\usepackage{amsthm}

\usepackage{graphicx}
\usepackage{diagbox} 

\usepackage{booktabs}

\usepackage{caption3} 
\DeclareCaptionOption{parskip}[]{} 
\usepackage[small]{caption}

\usepackage{geometry}
\usepackage{setspace,graphicx,multirow}
\usepackage{subcaption}
\usepackage{array}
\usepackage{algorithm}
\usepackage[]{algpseudocode}
\makeatletter
\def\BState{\State\hskip-\ALG@thistlm}
\makeatother
\usepackage{algcompatible}
\usepackage{url}

\usepackage{multirow}
\usepackage{hhline}
\usepackage{xspace}

\usepackage{color}

\ifodd 1
\newcommand{\com}[1]{\textbf{\color{blue} (COMMENT: #1)}} 
\else

\newcommand{\com}[1]{}
\fi

\usepackage{enumitem}
\setenumerate[1]{itemsep=0pt,partopsep=0pt,parsep=\parskip,topsep=0pt}
\setitemize[1]{itemsep=0pt,partopsep=0pt,parsep=\parskip,topsep=0pt}
\setdescription{itemsep=0pt,partopsep=0pt,parsep=\parskip,topsep=0pt}

\begin{document}
\raggedbottom
\bibliographystyle{IEEEtran}
\bstctlcite{IEEEexample:BSTcontrol}

\title{Neural Beam Field for Spatial Beam RSRP Prediction}

\author{Keqiang~Guo, Yuheng~Zhong, Xin~Tong, Jiangbin~Lyu,~\textit{Member,~IEEE}, and~Rui~Zhang,~\textit{Fellow,~IEEE}
\thanks{K. Guo, Y. Zhong, X. Tong, and J. Lyu are with the School of Informatics, Xiamen University (XMU), China, the Shenzhen Research Institute of XMU, China, and the Sichuan Institute of XMU, China. \textit{Corresponding author: Jiangbin Lyu} (email: ljb@xmu.edu.cn).

Rui Zhang is with the Department of Electrical and Computer Engineering, National University of Singapore, Singapore (email: \text{elezhang@nus.edu.sg}). 
}
}


\maketitle

\begin{abstract}
Accurately predicting beam-level reference signal received power (RSRP) is essential for beam management in dense multi-user wireless networks, yet challenging due to high measurement overhead and fast channel variations. This paper proposes Neural Beam Field (NBF), a hybrid neural-physical framework for efficient and interpretable spatial beam RSRP prediction. Central to our approach is the introduction of the Multi-path Conditional Power Profile (MCPP), a learnable physical intermediary representing the site-specific propagation environment. This approach decouples the environment from specific antenna/beam configurations, which helps the model learn site-specific multipath features and enhances its generalization capability. We adopt a decoupled ``blackbox-whitebox" design: a Transformer-based deep neural network (DNN) learns the MCPP from sparse user measurements and positions, while a physics-inspired module analytically infers beam RSRP statistics. To improve convergence and adaptivity, we further introduce a Pretrain-and-Calibrate (PaC) strategy that leverages ray-tracing priors for physics-grounded pretraining and then RSRP data for on-site calibration. Extensive simulation results demonstrate that NBF significantly outperforms conventional table-based channel knowledge maps (CKMs) and pure blackbox DNNs in prediction accuracy, training efficiency, and generalization, while maintaining a compact model size. The proposed framework offers a scalable and physically grounded solution for intelligent beam management in next-generation dense wireless networks.

\begin{IEEEkeywords}
Neural Beam Field, Multipath Conditional Power Profile, Channel Knowledge Map, Beam‐level RSRP, Transformer 
\end{IEEEkeywords}

\end{abstract}

%
\section{Introduction}

With the continuous densification of mobile communication networks, network operators face unprecedented challenges in maintaining scalable, highly spectral-efficient and reliable service coverage. Although massive Multiple-Input Multiple-Output (MIMO) and millimeter-wave technologies promise significant improvements in capacity and user throughput, the practical deployment of dense wireless networks is significantly constrained by the difficulty of timely acquiring Channel State Information (CSI) under realistic multi-user scenarios\cite{castaneda_overview_2017}. 
Traditionally, accurate CSI acquisition involves comprehensive measurements at high spatial-temporal-spectral resolutions, which inevitably leads to prohibitive overheads and excessive latency, severely limiting network scalability and real-time applicability \cite{li_beam_2020}.

Traditional channel estimation for MIMO beamforming mainly relies on two methods, i.e., pilot-based CSI estimation and beam scanning. The former method provides full CSI but incurs high overhead and latency, with the pilots given in the fifth-generation (5G) standards in the form of downlink CSI-RS or uplink SRS signals\cite{li_beam_2020}.
In comparison, the beam scanning method selects beam pairs from a codebook to estimate Reference Signal Received Power (RSRP) or Signal-to-Interference-plus-Noise Ratio (SINR) for user-specific beam pair links (BPLs).
To alleviate the burden of CSI acquisition, reduced-dimensional channel indicators such as RSRP or SINR have been widely adopted in practical network deployments \cite{zhang_physics_2024}. 
Nevertheless, even these simplified RSRP-based measurements can hardly satisfy real-time requirements due to varying propagation conditions and user mobility. Moreover, dense network conditions exaggerate the complexity, as complete measurement of RSRP for all user-beam combinations often exceeds channel coherence time, rendering CSI outdated before it can be utilized for resource allocation or beam scheduling \cite{ichkov_hbf_2025}. 
Moreover, the beam-specific RSRP is highly directional and thus relies on the site-specific blockages/shadowing/multi-path conditions.
These critical issues demand innovative approaches capable of predicting spatial RSRP fields under diverse beamforming configurations and environmental conditions.

Recent developments propose various spatial interpolation and estimation techniques to overcome CSI acquisition challenges by constructing radio map\cite{bi_engineering_2019} or channel knowledge map (CKM)\cite{zeng_tutorial_2024} from limited spatial samples. Early methods commonly assume spatial smoothness of received power, employing techniques such as Kriging \cite{boccolini_wireless_2012}, thin-plate spline interpolation \cite{bazerque_group_2011}, kernel methods \cite{romero_learning_2017}, and Gaussian radial basis functions \cite{hamid_nonparametric_2017}. Alternative methods, including sparse representation \cite{bazerque_distributed_2010} and low-rank matrix or tensor completion \cite{zhang_spectrum_2020}, have also validated theoretical feasibility under structural priors like sparsity or low-rankness. A new scatterer-based channel model is also proposed in \cite{Sun_machine_2023} for long-scale channel gain map estimation.
However, these methods might suffer from notable performance degradation in complex urban environments characterized by dynamic multipath propagation, blockage effects, and positioning uncertainties\cite{bi_engineering_2019,zeng_tutorial_2024}. 

In contrast, deep learning methods have shown substantial promise in capturing intricate spatial-temporal correlations and providing robust predictions, offering new opportunities for accurate prediction of RSRP \cite{zhao_nerf2_2023,lu_newrf_2024,orekondy_winert_2022}.
Inspired by Neural Radiance Fields (NeRF) \cite{mildenhall_nerf_2020} from computer vision, recent works further explore deep neural network (DNN)-based spatial field estimation methods integrated with physical propagation mechanisms. 
Notably, Neural RF Radiation Fields (NeRF2) \cite{zhao_nerf2_2023} predict signal features at arbitrary locations from sparse measurements, while Neural Wireless Radiation Fields (NeWRF) explicitly incorporate electromagnetic propagation physics into neural frameworks to enhance modeling accuracy in heterogeneous environments \cite{lu_newrf_2024}. Additionally, WiNeRT \cite{orekondy_winert_2022} leverages DNNs to model intricate interactions between rays and reflective surfaces, capturing complex multipath propagation dynamics.
Nevertheless, these approaches have yet to explicitly incorporate beam-specific RSRP measurements in practical 5G deployments.


Motivated by these limitations, we introduce the \textit{Neural Beam Field (NBF)}, a novel DNN-based method tailored for predicting spatial beam RSRP statistics, explicitly incorporating directional beamforming and multipath propagation physics.  
Our main contributions are summarized as follows:
\begin{itemize}
    \item First, we introduce the new notion of \textit{Multi-path Conditional Power Profile (MCPP)}, which serves as an essential intermediary between physical wave propagation in a site-specific environment and the add-on antenna panel/beam configurations. Based on MCPP, we further derive analytical formulas for beam RSRP statistics, which facilitate easier learning and generalization under diverse settings. Therefore, the incorporation of beam RSRP and MCPP-based physical grounding provides one key innovation compared with existing neural field methods.





    \item Second, we propose a decoupled ``\textit{blackbox-whitebox}" approach to construct NBF, by 1) learning the site-specific MCPPs using a tailored design of Transformer-based DNN, and 2) inferring the beam RSRP statistics based on analytical formulas. This ``whitebox" analytical modeling combined with ``blackbox" neural representations results in an interpretable and accurate predictive framework. 
    

    \item Finally, we further propose a \textit{Pretrain-and-Calibrate (PaC)} strategy in cases when prior information about MCPPs are available. Leveraging on 1) tailored MCPP loss functions that could handle un-ordered/varying number of paths and 2) regularized calibration loss that preserves the model's physical groundings and enhances finetuning stability, the PaC strategy helps tame the NBF into a robust locally optimal point that could then adapt to unknown/random factors by on-site RSRP measurements.
   

    \end{itemize}
    
Numerical results validate the analytical formulas through extensive Monte Carlo (MC) simulations. Moreover, it is shown that the proposed approach significantly outperforms conventional table-based CKMs and pure blackbox neural models in terms of prediction accuracy, storage efficiency, and generalization capability.
Our proposed NBF method for spatial beam RSRP prediction holds potential in intelligent beam management and user scheduling in dense wireless networks.

\textit{Notations}: 
Symbols for vectors (lower case) and matrices (upper case) are in boldface. 
$\otimes$ ($\odot$) represents the Kronecker (Hadamard) product. 
$(\cdot)^T$ and $(\cdot)^*$ represent the transpose and conjugate, respectively. 
\text{Re}($\cdot$) takes the real part of a complex number while $\angle \cdot$ takes the angle of a complex number.

\section{System Model}\label{SectionModel}

\begin{figure}[t]
  \centering
  \includegraphics[width=0.75\linewidth,  trim=0 0 0 0,clip]{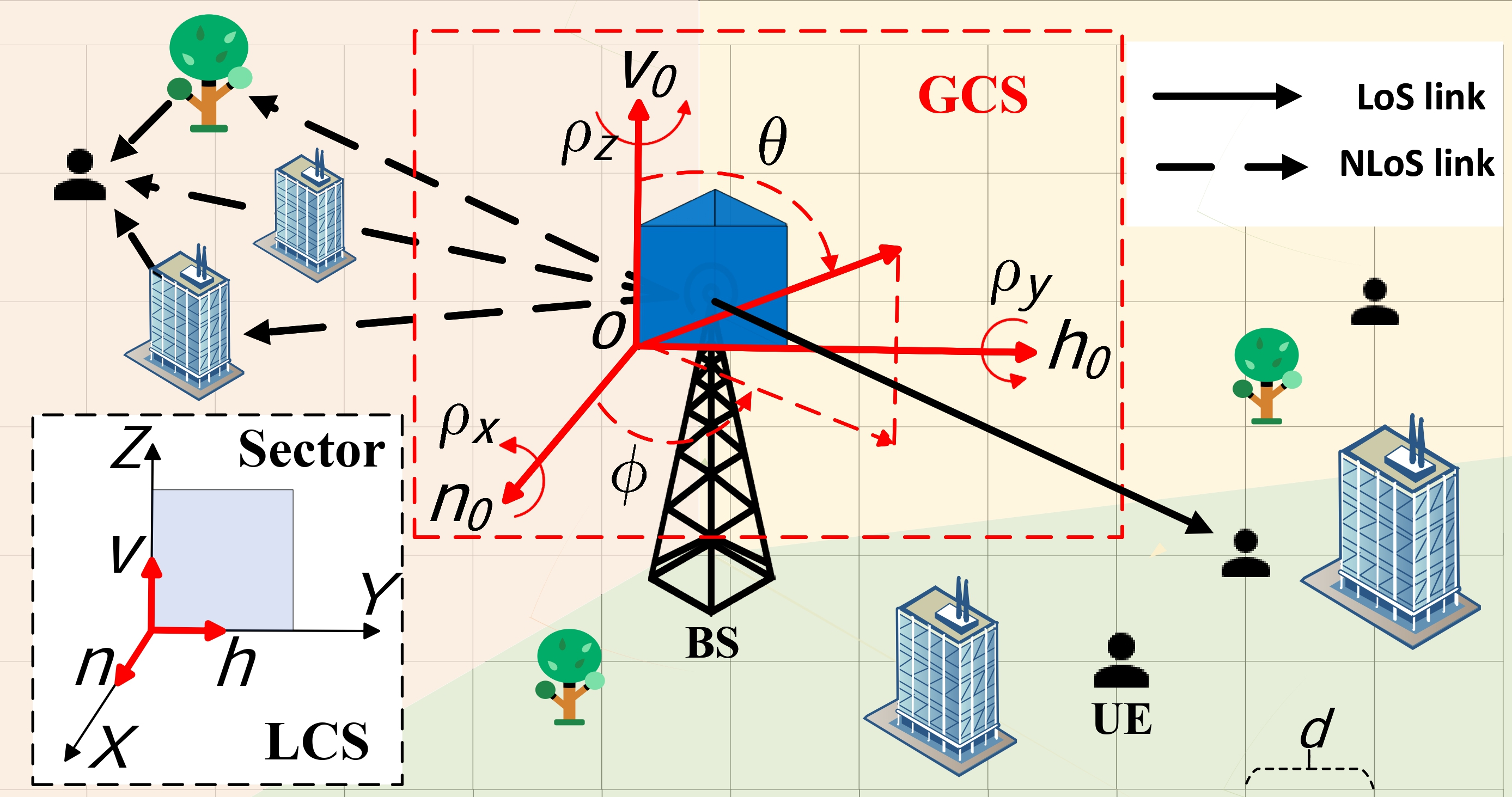}
  \caption{%
    Three‐sector urban macrocell scenario with site-specific blockages and multipath propagation conditions.
  }
  \label{fig:scenario}
\end{figure}

As shown in Fig.~\ref{fig:scenario}, we consider downlink communications in a typical three‐sector urban macrocell scenario with site-specific blockages and multipath propagation conditions, where the service region $\mathcal{A}$ is partitioned into regular grids with spacing \(d\) meters (m). 
For simplicity, the user equipments (UEs) are assumed to be equipped with a single isotropic antenna at height $h_\textrm{UE}$ m.
The base station (BS) is located at a known position with height \(h_\textrm{BS}\) m, and employs three sector antenna panels each configured as a uniform planar array (UPA) of size \(N_\text{tx}\triangleq N_\text{tx,y}\times N_\text{tx,z}\). 
Each antenna element is assumed to exhibit a certain element radiation pattern (ERP), e.g., the directional antenna model as in 3GPP TR 38.901\cite{3GPPChannelModel38901}, denoted by $G\left(\Pi_\text{tx}\right)$, which represents the power pattern as seen from the azimuth and elevation angles $\Pi_\text{tx} \triangleq \left(\theta_\text{tx}, \varphi_\text{tx}\right)$ in the global coordinate system (GCS), whose origin co-locates with the reference BS antenna element.



\subsection{Antenna Panel Rotation and Array Response Vector}

Each antenna panel $i\in\mathcal{I}=\{1,2,3\}$ has counterclockwise rotation angles $\rho_{\text{x}, i}$, $\rho_{\text{y}, i}$ and $\rho_{\text{z}, i}$ about the $x$, $y$, and $z$ axes, respectively, based from the default Y-O-Z plane in the GCS.
Let \(\mathbf{n}_0\), \(\mathbf{h}_0\) and \(\mathbf{v}_0\) denote the GCS orthonormal basis vectors corresponding to the normal, horizontal and vertical directions, respectively.
For a given panel, when the array is rotated by angles \(\rho_\text{x}\), \(\rho_\text{y}\) and \(\rho_\text{z}\) about the $x$, $y$, and $z$ axes in sequence, the resulting composite rotation matrix is given by \cite{3GPPChannelModel38901}
\begin{equation}
    \mathbf{R} = \mathbf{R}_\text{z}(\rho_\text{z})\,\mathbf{R}_\text{y}(\rho_\text{y})\,\mathbf{R}_\text{x}(\rho_\text{x}).
\end{equation}
As a result, the basis vectors after rotation are given by
\begin{equation}
    \mathbf{n} = \mathbf{R}\,\mathbf{n}_0,\quad
    \mathbf{h} = \mathbf{R}\,\mathbf{h}_0,\quad
    \mathbf{v} = \mathbf{R}\,\mathbf{v}_0,
\end{equation}
which form the orthonormal basis for the panel's local coordinate system (LCS), as shown in Fig.~\ref{fig:scenario}.

Denote $\boldsymbol{u}_\text{tx} \triangleq [\cos\varphi_\text{tx} \sin\theta_\text{tx},\sin\varphi_\text{tx} \sin\theta_\text{tx},\cos\theta_\text{tx}]^T$ as the 3-dimensional (D) unit vector of a given wave propagation path w.r.t. the origin, with $\Pi_\text{tx} \triangleq (\varphi_\text{tx}, \theta_\text{tx})$ denoting the corresponding (horizontal, vertical) direction of departure (DOD) angle in the GCS.
Consider one typical antenna panel which locates in the Y-O-Z plane in the GCS.
The components of its array response vector (ARV) along the $y\text{-}$ and $z\text{-}$axis are given by
\begin{align} 
\boldsymbol{a}_\text{tx,y}(\Pi_\text{tx}) \triangleq [1, \cdots, e^{j\frac{2\pi f_\text{c}}{c} (\boldsymbol{r}_{N_\text{tx,y}}^T \boldsymbol{u}_\text{tx})}]^T \in \mathbb{C}^{N_\text{tx,y} \times 1}, \label{ay}\\
\boldsymbol{a}_\text{tx,z}(\Pi_\text{tx}) \triangleq [1, \cdots, e^{j\frac{2\pi f_\text{c}}{c} (\boldsymbol{r}_{N_\text{tx,z}}^T \boldsymbol{u}_\text{tx})}]^T \in \mathbb{C}^{N_\text{tx,z} \times 1},\label{az}
\end{align}%
where $\boldsymbol{r}_{n_{\text{tx,y}}} \triangleq (n_{\text{tx,y}}-1)d_{\text{y}} \mathbf{h}$ and $\boldsymbol{r}_{n_{\text{tx,z}}} \triangleq (n_{\text{tx,z}}-1)d_{\text{z}} \mathbf{v}$ denote the position vectors of the $(n_{\text{tx,y}},n_{\text{tx,z}})$-th element along the $y\text{-}$ and $z\text{-}$axis, with $d_{\text{y}}$ and $d_{\text{z}}$ being the element spacing, respectively;
$c$ denotes the speed of light and $f_\text{c}$ denotes the carrier frequency.
Then, the ARV is given by $\boldsymbol{a}_\text{tx}(\Pi_\text{tx}) \triangleq \boldsymbol{a}_{\text{tx,y}}(\Pi_\text{tx}) \otimes \boldsymbol{a}_{\text{tx,z}}(\Pi_\text{tx})\in \mathbb{C}^{N_\text{tx} \times 1}$.
The ARV for other panels with different rotations can be expressed in similar manners.

\subsection{Channel Model}

For the pure propagation channel (assuming an isotropic antenna at both sides) between the BS and the target UE, assume that there exists $L$ significant propagation paths. Each path $l\in\mathcal{L}\triangleq\{1,\cdots,L\}$ is associated with unit DOD vector $\boldsymbol{u}_{\text{tx},l}$ and unit direction of arrival (DOA) vector $\boldsymbol{u}_{\text{rx},l} \triangleq [\cos\varphi_{\text{rx},l} \sin\theta_{\text{rx},l},\sin\varphi_{\text{rx},l} \sin\theta_{\text{rx},l},\cos\theta_{\text{rx},l}]^T$, with $\Pi_{\text{rx},l} \triangleq (\varphi_{\text{rx},l}, \theta_{\text{rx},l})$ denoting the corresponding (horizontal, vertical) DOA angle w.r.t. the UE position.
As a result, the BS-UE channel for the $l$-th path is given by
\begin{equation}
\boldsymbol{g}_l\triangleq \textstyle\sqrt{G(\Pi_{\text{tx},l})} \alpha_l e^{j\Phi_{l}} e^{-j2\pi f_\text{c}\tau_{l}} \boldsymbol{a}_\text{tx}(\Pi_{\text{tx},l}),
\end{equation}
where $\alpha_l$, $\Phi_{l}$ and $\tau_{l}$ denote the amplitude, phase and propagation delay of the $l$-th path, respectively.
Note that for a given UE position in a quasi-static environment, the per-path geometric parameters including $\Pi_{\text{tx},l}$, $\Pi_{\text{rx},l}$, $\alpha_l$ and $\tau_{l}$ mainly depend on the environment geometry and could be obtained from ray tracing or site survey and stored in the form of CKMs. On the other hand, the channel phase $\Phi_{l}$ is affected by the micro user movements/environment changes, which is thus treated as a random variable. 

Therefore, we introduce the notion of \textit{Single-path Conditional Power (SCP)} for each path $l\in\mathcal{L}$, denoted as
\begin{equation}\label{SCP}
    \mathrm{SCP}_l=\alpha_l^2\triangleq  p_l\bigl(\boldsymbol{u}_{\text{tx},l},\,\boldsymbol{u}_{\text{rx},l},\,\tau_l\bigr),
\end{equation}%
where $(\boldsymbol{u}_{\text{tx},l},\,\boldsymbol{u}_{\text{rx},l},\,\tau_l\bigr)$ collectively represents the 7-D path condition. Note that we choose the unit direction vectors $\boldsymbol{u}_{\text{tx},l}$ and $\boldsymbol{u}_{\text{rx},l}$ instead of the angles $\Pi_{\text{tx},l}$ and $\Pi_{\text{rx},l}$ for better representation and easier learning of periodic angle features.
Furthermore, we introduce the new notion of \textit{Multi-path Conditional Power Profile (MCPP)} for all paths in $\mathcal{L}$, denoted as
\begin{equation}\label{MCPP}
    \mathrm{MCPP} \triangleq \{\mathrm{SCP}_1,\,\mathrm{SCP}_2,\,\dots,\,\mathrm{SCP}_L\},
\end{equation}
which is a set of $L$ power entries in the 7-D conditional space.

\textbf{Remark:} The MCPP for the pure propagation channel serves as an essential intermediary between physical wave propagation in a site-specific environment and the add-on antenna/beam configurations, which jointly determine the beam RSRP statistics. 
Therefore, we propose a decoupled ``blackbox-whitebox" approach by 1) learning the site-specific/irregular MCPPs using DNNs, and 2) inferring the beam RSRP statistics based on our derived analytical formulas in the following.

\subsection{Beam RSRP Statistics under Given Multipath Conditions}
\subsubsection{RSRP under Given Beamforming Scheme}
For the purpose of exposition, here we consider analog beamforming with discrete Fourier transform (DFT) codebooks.\footnote{Note that our proposed method is general and can be extended to other beamforming schemes such as hybrid digital/analog beamforming.}
Specifically, for the typical UPA panel in the Y-O-Z plane, the components of its beamforming vector
along the $y\text{-}$ and $z\text{-}$axis are given by
\begin{align} 
\boldsymbol{w}_\text{tx,y}(\xi_\text{tx,y}) \triangleq [1, e^{j \xi_\text{tx,y}},\cdots, e^{j (N_\text{tx,y}-1) \xi_\text{tx,y}}]^T \in \mathbb{C}^{N_\text{tx,y} \times 1}, \label{wy}\\
\boldsymbol{w}_\text{tx,z}(\xi_\text{tx,z}) \triangleq [1, e^{j \xi_\text{tx,z}},\cdots, e^{j (N_\text{tx,z}-1) \xi_\text{tx,z}}]^T \in \mathbb{C}^{N_\text{tx,z} \times 1},\label{wz}
\end{align}%
where $\xi_\text{tx,y}$, $\xi_\text{tx,z}\in [-\pi,\pi)$ denote the spatial frequencies along the $y\text{-}$ and $z\text{-}$axis, respectively.
Then, the overall DFT-type beamforming vector with normalized power is given by 
\begin{equation}
    \boldsymbol{w}_\text{tx}(\xi_\text{tx,y},\xi_\text{tx,z}) \triangleq \frac{1}{\sqrt{N_\text{tx}}} \boldsymbol{w}_{\text{tx,y}}(\xi_\text{tx,y}) \otimes \boldsymbol{w}_{\text{tx,z}}(\xi_\text{tx,z}).
\end{equation}

In the downlink, the received signal is then given by
\begin{equation}
y\triangleq \textstyle\sum\nolimits_{l\in\mathcal{L}}\boldsymbol{g}_{l}^T\,\mathbf{w}_\text{tx}\,s
  \;+\;n,
\end{equation}
where \(s\) denotes the transmit symbol with power $P_\text{t}$ Watt (W) and $n$ denotes the additive white Gaussian noise.
The beam RSRP normalized to $P_\text{t}$ is then given by
\begin{equation}\label{RSRPdef}
    P_\text{r}(\mathrm{MCPP},\mathbf{w}_\text{tx})\triangleq |\textstyle\sum\nolimits_{l\in\mathcal{L}}\boldsymbol{g}_{l}^T\,\mathbf{w}_\text{tx}|^2,
\end{equation}
with $P_\text{r,dB}$ denoting the corresponding normalized RSRP in dB.


\subsubsection{Beam RSRP Statistics under Given MCPP}
Here we derive the beam RSRP statistics under given MCPP.
For simplicity, we first consider the narrowband scenario while leaving the wideband case for future extension. 
The results are summarized in the following proposition.
\newtheorem{prop}{Proposition}
\begin{prop}\label{Prop_Beamforming}
For independent random channel phases $\Phi_{l}$, $l\in\mathcal{L}$ uniformly distributed in $[0,2\pi)$, the mean and variance of the normalized beam RSRP in \eqref{RSRPdef} are given by
\begin{equation}
    \mu_{\mathrm{RSRP}}(\mathrm{MCPP},\mathbf{w}_\text{tx})
    = \textstyle\sum\nolimits_{l\in\mathcal{L}} \gamma_l,
\end{equation}
\begin{equation}
    \sigma_{\mathrm{RSRP}}^{2}(\mathrm{MCPP},\mathbf{w}_\text{tx})
    = \left(\textstyle\sum\nolimits_{l\in\mathcal{L}} \gamma_l\right)^2 - \textstyle\sum\nolimits_{l\in\mathcal{L}} \gamma_l^2,
\end{equation}%
where $\gamma_l\triangleq G(\Pi_{\text{tx},l}) p_l |\Delta_l|^2$ denotes the average power gain, and $|\Delta_l|^2= \frac{1}{N_\text{tx}} \big| S_{N_{\text{tx,y}}}(\zeta_{\text{tx,y},l} +\xi_{\text{tx,y}}) \big|^2 \big| S_{N_{\text{tx,z}}}(\zeta_{\text{tx,z},l} +\xi_{\text{tx,z}})\big|^2$ denotes the power gain of the array factor $\Delta_l \triangleq \boldsymbol{a}_\text{tx}^T(\Pi_{\text{tx},l})\,\mathbf{w}_\text{tx}$, for path $l\in\mathcal{L}$, with $\zeta_{\text{tx,y},l}\triangleq \frac{2\pi f_\text{c} d_{\text{y}}}{c} \mathbf{h}^T \boldsymbol{u}_\text{tx}(\Pi_{\text{tx},l})$, $\zeta_{\text{tx,z},l}\triangleq \frac{2\pi f_\text{c} d_{\text{z}}}{c} \mathbf{v}^T \boldsymbol{u}_\text{tx}(\Pi_{\text{tx},l})$ and $S_N(\cdot)$ given by \eqref{SN}.
\end{prop}
\textit{Proof:} Please refer to Appendix A.$\blacksquare$

\subsection{Practical Collection/Synthetic Generation of RSRP Data}

In current 5G standards, beam RSRP could be obtained from pilot signals such as synchronization signal blocks (SSBs) and CSI-reference signal (CSI‐RS). 
These measurement results could be combined with UE location information to construct a CKM for spatial beam RSRP prediction.
In practical scenarios, however, the RSRP measurements might be performed by different UEs at different time and locations with possible positioning errors, which requires effective ways for data aggregation. 
This paper proposes a grid‐based weighted aggregation method for obtaining stable RSRP statistics. 
Specifically, the RSRP measurements are first mapped to neighboring grid anchors based on the estimated UE position and its confidence interval, with normalized sample weights based on, e.g., inverse distance weighting (IDW). 
Furthermore, for timely update of the CKM, a temporal weight decay mechanism could be introduced by gradually decaying the sample weights over time. 
As a result, each grid anchor absorbs spatial/temporal RSRP measurements and accumulates stable RSRP statistics.


On the other hand, as a preliminary study, we also provide ways for synthetic generation of RSRP data. In compliance with the 3GPP TR38.901 hybrid channel model\cite{3GPPChannelModel38901}, we employ SionnaRT\cite{hoydis_sionna_2023} to generate site-specific MCPPs for a given city region based on ray tracing, while utilizing QuaDRiGa\cite{hhi_quadriga_manual_2023} to generate spatially consistent random perturbations for the MCPPs to simulate unknown/random environmental factors. The final MCPP at each location is then obtained as the weighted sum of the above deterministic and random MCPP components, with normalized weight $0\leq\beta<1$ for the latter.

\section{Design and Calibration of Neural Beam Field}\label{SectionUserPerformance}
The objective of NBF is to predict the beam RSRP statistics under any given UE location $\mathbf{x}$ and beam configuration $\mathbf{w}_\text{tx}$.
Based on the spatially distributed RSRP measurements, conventional table-based CKMs could be employed by storing the beam RSRP statistics in discrete map grids and performing interpolation in under-sampled locations/beams.
However, it typically incurs large storage overhead to store the complete set of CKMs under various configuration conditions, while remaining challenging for accurate spatial interpolation/nonlinear inference. 
On the other hand, DNN-based blackbox models could be employed to learn the complicated nonlinear mappings between UE positions/beam configurations and the corresponding RSRP statistics.
Nevertheless, pure blackbox models lack physical interpretation and typically require sufficiently sampled datasets from the high dimensional input space, thereby often associated with prolonged training time and difficulty of generalization beyond the training dataset.

In this section, we propose a decoupled ``blackbox-whitebox" approach to construct the NBF, by 1) designing a tailored DNN based on Transformer\cite{viT} to learn the site-specific/irregular MCPPs, and 2) inferring the beam RSRP statistics based on our derived analytical formulas in \textbf{Proposition 1}.
Such an approach leverages on the essential role of MCPP in amalgamating propagation channels and beam configurations, which thus enables easier learning and generalization.

\begin{figure}[t]
  \centering
  \includegraphics[width=1\linewidth]{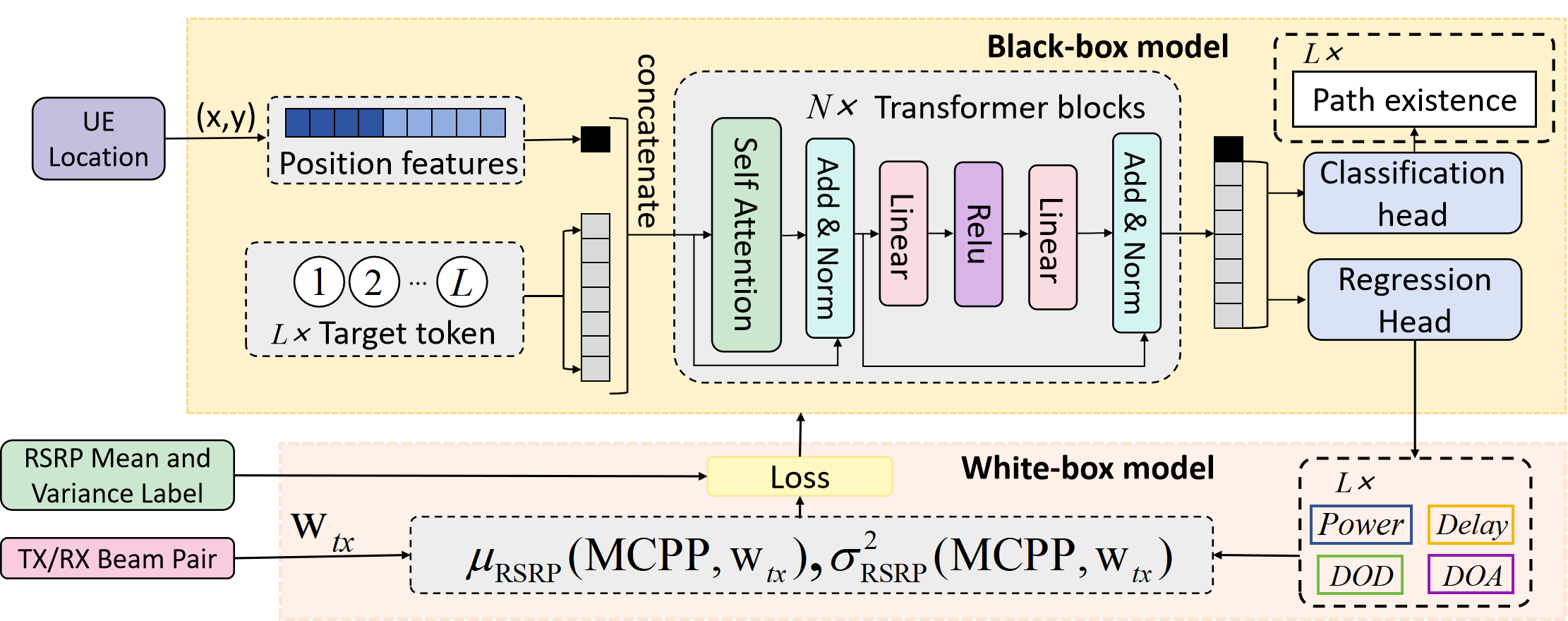}
  \caption{NBF Network Architecture.}
  \label{fig:Network_Architecture}
\end{figure}

\subsection{Neural Beam Field Design}
The overall NBF design is illustrated in Fig.~2, which consists of 1) a Transformer-based blackbox model $f_\vartheta (\mathbf{x})$ that predicts the MCPP at a given UE location $\mathbf{x}$, and 2) a whitebox model that maps to beam RSRP statistics based on \textbf{Proposition 1}. 
For the purpose of exposition, we consider the mean RSRP as the output while higher order statistics could be treated in similar manners.
Therefore, the NBF function is given by
\begin{equation}
\mathrm{NBF}(\mathbf{x},\mathbf{w}_\text{tx})
\triangleq \mu_{\mathrm{RSRP}}\big(f_\vartheta(\mathbf{x}),\mathbf{w}_\text{tx}\big),
\end{equation}%
with UE location $\mathbf{x}$ and beam configuration $\mathbf{w}_\text{tx}$ as the input.
In the following, we focus on the DNN design.


Firstly, for the input of UE location $\mathbf{x}=(x,y)$, we adopt the random Fourier position embedding\cite{kirillov_segment_2023} to obtain a $d_{\text{model}}$-D feature vector $\mathbf{X}_\text{UE}\in\mathbb{R}^{1\times d_{\text{model}}}$, which can then be treated as a token for subsequent Transformer blocks.
In addition, in order to output the MCPP which consists of $L$ SCPs, we also place $L$ learnable target tokens $\mathbf{X}_\text{target}\in\mathbb{R}^{L\times d_{\text{model}}}$ at the input.
Secondly, the concatenated input tokens $\mathbf{X}\triangleq[\mathbf{X}_\text{UE};\mathbf{X}_\text{target}]$ are processed by the Transformer encoder, comprising $N$ blocks of multi-head self-attention (MSA), multi-layer perceptron (MLP), layer normalization (LN), and residual connections\cite{viT}. 
Notably, 
the learnable target tokens attend to the UE location token along the blocks to extract key information related to location-specific MCPPs.
Finally, each of the target tokens processed by Transformer is fed to two MLPs, i.e., 1) a two-layer regression head that produces 8 features corresponding to one SCP in \eqref{SCP}, thereby composing the predicted MCPP $\hat{\mathbf{Y}}=f_\vartheta(\mathbf{x})$; and 2) a one-layer classification head with output $a_l\in \{0,1\}$ that tells the existence of path $l$ (in cases with fewer than $L$ paths).

\subsection{End-to-End Learning Based on RSRP Measurements}\label{SectionE2E}
Based on the above NBF design, we then train the DNN such that the end-to-end (blackbox-whitebox) prediction of beam-specific mean RSRP matches as close as possible to the corresponding labels from measurement data.
Note that the whitebox module, which implements the analytical formulas from \textbf{Proposition 1}, contains no trainable parameters. It serves as a fixed, differentiable mapping function that connects the MCPP to the mean RSRP. This structure allows for end-to-end gradient-based optimization of the blackbox part.
Finally, the Smooth-$L_1$ function is chosen as the regression loss $\mathcal{L}_{\text{RSRP}}$ because it is less sensitive to outliers than the $L_2$ loss. The gradients from this loss back-propagate through the whitebox module to update only the blackbox parameters $\vartheta$.

\subsection{MCPP-Prior Aided Pretraining and On-Site Calibration}
Due to the highly nonlinear mapping task in NBF, the training process might trap in undesired local points. 
To achieve accelerated convergence and better optimality, we further propose a \textit{Pretrain-and-Calibrate (PaC)} strategy in cases when prior (but not necessarily exact) information about MCPPs are available, e.g., through ray tracing in an environment with known geometry and materials.

In the pretraining phase, MCPP labels in the form of \eqref{MCPP} are available, denoted as 
$\bar{\mathbf{Y}}$. 
We aim to minimize the difference between the predicted MCPP $\hat{\mathbf{Y}}$ and ground-truth MCPP $\bar{\mathbf{Y}}$.
However, two issues exist, i.e., 1) some labels $\bar{\mathbf{Y}}$ might have paths fewer or more than $L$, 
and 2) the permutation-invariant path indices require per-path matching between $\hat{\mathbf{Y}}$ and $\bar{\mathbf{Y}}$.
Regarding the first issue, note that $L$ could be chosen as the typical (maximum) number of significant paths under given environment type and carrier frequency. In some outlier locations with more than $L$ paths, clustering methods such as weighted K-means could be employed to merge them into $L$ significant paths. On the other hand, for locations with fewer than $L$ paths, we utilize the path existence prediction $a_l$ along with binary cross-entropy (BCE) loss to learn the situation.

To handle un-ordered/varying number of paths, we propose a \textit{set prediction} loss based on bipartite matching. For each training sample, we first find an optimal one-to-one assignment between the $L$ predicted paths and the ground-truth paths using the Hungarian algorithm. The assignment cost matrix considers both the regression difference of path parameters and the classification score for path existence.
After matching, the total loss for the pre-training stage, $\mathcal{L}_\text{pretrain}$, is the weighted sum of a classification loss $\mathcal{L}_\text{cls}$ and a regression loss $\mathcal{L}_\text{reg}$, i.e.,
\begin{equation}
\mathcal{L}_\text{pretrain} \triangleq \mathcal{L}_\text{cls} + \lambda_\text{reg} \mathcal{L}_\text{reg},
\end{equation}%
where $\mathcal{L}_\text{cls}$ adopts a binary cross-entropy (BCE) loss computed over all $L$ predicted path existence logits, and $\mathcal{L}_\text{reg}$ adopts the Smooth-$L_1$ loss computed only between the parameters of the matched pairs, with the weight $\lambda_\text{reg}$ to balance the two losses. 
During the calibration phase, besides the loss $\mathcal{L}_{\text{RSRP}}$ for mean RSRP prediction, to enhance the stability of fine-tuning and mitigate potential catastrophic forgetting, we introduce a regularization loss $\lambda_{\text{feat}}$ with weight $\lambda_{\text{feat}}$ based on feature representation alignment, resulting in a joint loss function, i.e.,
\begin{equation}
\mathcal{L}_{\text{finetune}} \triangleq \mathcal{L}_{\text{RSRP}} + \lambda_{\text{feat}} \mathcal{L}_{\text{feat}}.
\end{equation}
More specifically, the parameters $\vartheta$ of the pretrained network are loaded and frozen to serve as a \textit{reference model}. The term $\mathcal{L}_{\text{feat}}$ then quantifies the mean squared error (MSE) between the intermediate feature representations produced by the current model and those produced by the reference model. 
This regularized calibration approach effectively transfers the inductive bias acquired during pretraining to the fine-tuning stage, thus preserving the model's physical groundings while efficiently adapting to the unknown environmental factors. 
\subsection{Computational Complexity Analysis}
During forward inference, the NBF invokes a blackbox DNN with $L+1$ tokens passed through $N$ Transformer blocks to predict $L$ SCPs, which are then fed to the whitebox to map into mean RSRP.
According to \cite{vaswani_attention_2017}, the inference complexity is $O\big(N \cdot [(L+1)^2 \cdot d_{\text{model}} + (L+1)\cdot d_{\text{model}}^2]\big)$, with $d_{\text{model}}$ being the number of dimensions in each token.
In addition, the whitebox module processes each SCP independently and thus is of complexity $O(L)$.
In comparison, the training stage involves the same complexity order as in the inference stage\cite{CNNtime}, with the additional cost for loss evaluation. In particular, the MCPP path matching adopts the classic Hungarian algorithm which is of complexity $O(L^3)$.
Finally, we adopt the Adam optimizer which is guaranteed to converge to at least one locally optimal point\cite{TomLuoAdam}.
Also note that, whether training or inference, the computational complexity of NBF is independent of the geographical area size, which affects only the value range of UE locations and the number of available samples.


\section{Numerical Results}

In this section, we first verify \textbf{Proposition 1} through extensive MC simulations, and then demonstrate the effectiveness of NBF compared with representative benchmark schemes.
As discussed in Section I, RSRP prediction based on spatially distributed RSRP measurements mainly includes classic spatial interpolation methods\cite{boccolini_wireless_2012, bazerque_group_2011, romero_learning_2017, hamid_nonparametric_2017,bazerque_distributed_2010,zhang_spectrum_2020,Sun_machine_2023} and DNN-based methods\cite{zhao_nerf2_2023,lu_newrf_2024,orekondy_winert_2022}, which have yet to explicitly incorporate beam RSRP.
As a preliminary study, we consider inverse distance weighting (IDW) as a representative for the former and an MLP-based DNN for the latter, while leaving more exhaustive comparisons for future work.
IDW performs spatial interpolation from nearby grids\footnote{The neighbor range is empirically chosen as $3d$ for desired performance.}, which consists of two variants, i.e., 1) interpolating the mean RSRP labels directly, and 2) interpolating the MCPP first and then use the whitebox module for inference.
The MLP baseline performs end-to-end regression of mean RSRP, using normalized user coordinates and sin-cos encoded DFT beam spatial frequencies as inputs. 
Its model size is slightly larger than that of the NBF to ensure a fair comparison.
The following parameters are used: $h_\text{BS} = 20$ m, $h_\text{UE} = 1.5$ m, $L = 10$, $f_c = 3.5$ GHz, $N_{\mathrm{tx,y}} = 8$, $N_{\mathrm{tx,z}} = 4$, $\beta=0.5$, $\lambda_\text{reg}=5$, $\lambda_\text{feat}=0.1$, $d_{\text{model}}=256$, a square area with side length 256 m and $d=1$ m, a panel down-tilt angle of $15^\circ$, peak antenna element gain of 8 dBi, and half-wavelength inter-element spacing.


\subsection{Validation of Analytical Formulas}

\begin{figure}[ht]
  \centering
  \begin{subfigure}[b]{0.48\linewidth}
    \centering
    \includegraphics[width=\linewidth]{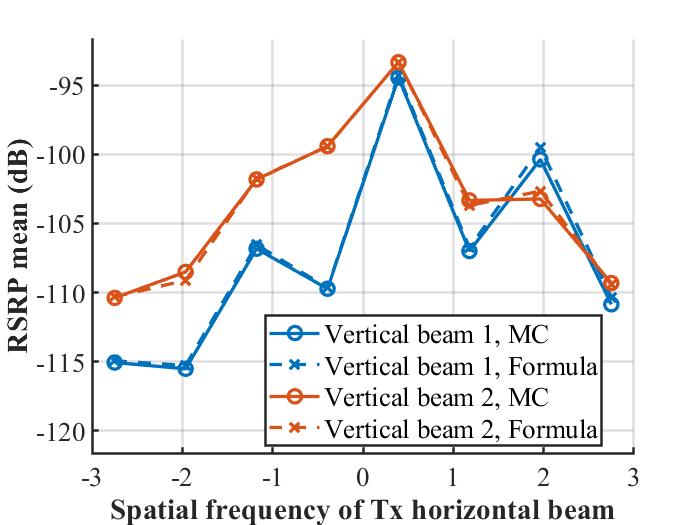} 
    \caption{}
    \label{fig:rsrp_mean_comparison}
  \end{subfigure}
  \hfill
  \begin{subfigure}[b]{0.48\linewidth}
    \centering
    \includegraphics[width=\linewidth]{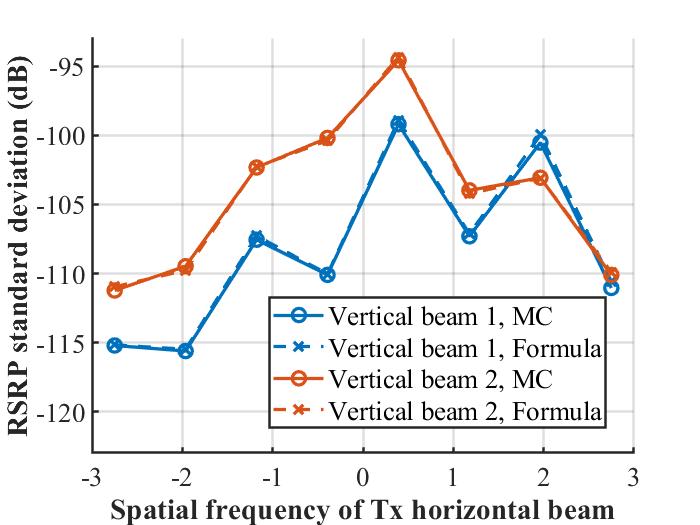}
    \caption{}
    \label{fig:rsrp_std_comparison}
  \end{subfigure}
  \caption{(a) Mean and (b) standard deviation of beam RSRP obtained by MC simulations and formulas under different beam configurations.}
  \label{fig:rsrp_mean_standard_comparison}
\end{figure}

We simulate an outdoor scenario using the QuaDRiGa toolbox, which implements 3GPP 38.901 channel models. In the MC simulations, we average over $N_{\mathrm{mc}}=200$ independent channel realizations, apply the DFT beamforming weights in each realization, and compute the empirical mean and standard deviation of the instantaneous received power.
For a given UE location, the results are plotted in Fig. \ref{fig:rsrp_mean_standard_comparison}, with good match with our analytical formulas across different beam configurations. 
Furthermore, good consistency has also been observed across the entire map, which is omitted due to space limit.




\subsection{Effectiveness of NBF}
For NBF training, we collect ray-traced user channels and RSRP samples on a site-specific map. 
Each sample consists of the UE position, antenna panel and beam configurations, together with the associated performance label. We use an $80/20$ split of dataset for training and validation, and optimize the Smooth-$L_{1}$ loss on the mean RSRP using the Adam optimizer and OneCycleLR learning rate schedule.

\begin{figure}[!htbp]
  \captionsetup{aboveskip=4pt, belowskip=2pt} 
  \centering
  \includegraphics[width=1.0\linewidth]{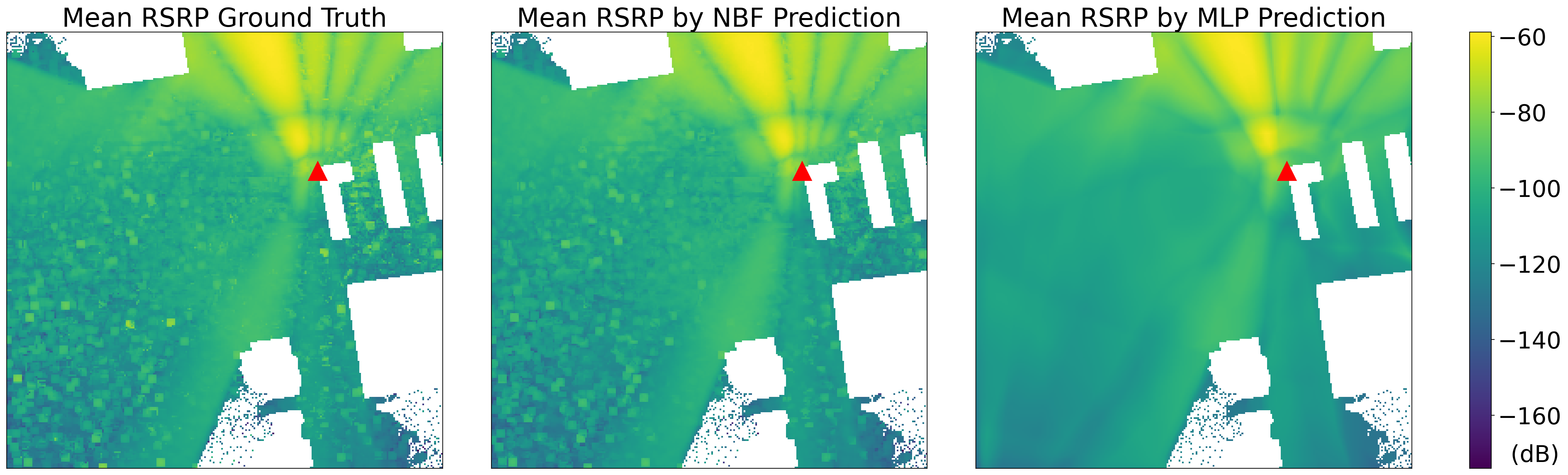}
  \caption{Spatial mean RSRP heatmap comparison for a given beam: ground truth (left), NBF (center), and MLP (right).}
  \label{fig:label_nbf_mlp_heatmap_comparison}
\end{figure}

Firstly, We evaluate the effectiveness of NBF through an illustrative example in Fig.~\ref{fig:label_nbf_mlp_heatmap_comparison}, which presents spatial mean RSRP heatmaps for a representative beam under ground truth, pretrained NBF, and MLP. It can be seen that the NBF prediction closely follows the ground-truth distribution and captures fine spatial details, while the MLP output appears overly smooth and misses local details.




Furthermore, quantitative results are compared in Table~\ref{tab:performance_comparison}, including mean absolute error (MAE) of mean RSRP prediction, model storage requirement in Mega-Bytes (MB), and computation/inference time in milliseconds (ms).
It can be seen that NBF provides the best overall performance. 
Although the RSRP-based IDW method is faster in inference, its prediction accuracy remains inferior to our NBF model even with larger storage (and correspondingly more RSRP measurements).


\begin{table}[ht]
    \centering
    \caption{Performance Comparison of NBF and Baseline Models}
    \label{tab:performance_comparison}
    \setlength{\tabcolsep}{3pt}
 \begin{footnotesize}   
    \begin{tabular}{lccc}
        \toprule 
        \textbf{Model} & \textbf{MAE (dB)} & \textbf{Storage (MB)} & \textbf{Time (ms)} \\
        \midrule
        \textbf{NBF (Proposed)} & 1.218 & 16.4 & 0.273 \\
        MLP (Baseline)     & 3.423 & 16.5 & 0.212 \\
        IDW (MCPP-based)   & 2.996 & 27.0 & 3.365 \\
        IDW (RSRP-based)   & 1.937 & 55.3 & 0.062 \\
        \bottomrule
    \end{tabular}
\end{footnotesize}
\end{table}

\begin{figure}[!htbp]
 \centering
 \includegraphics[width=\linewidth]{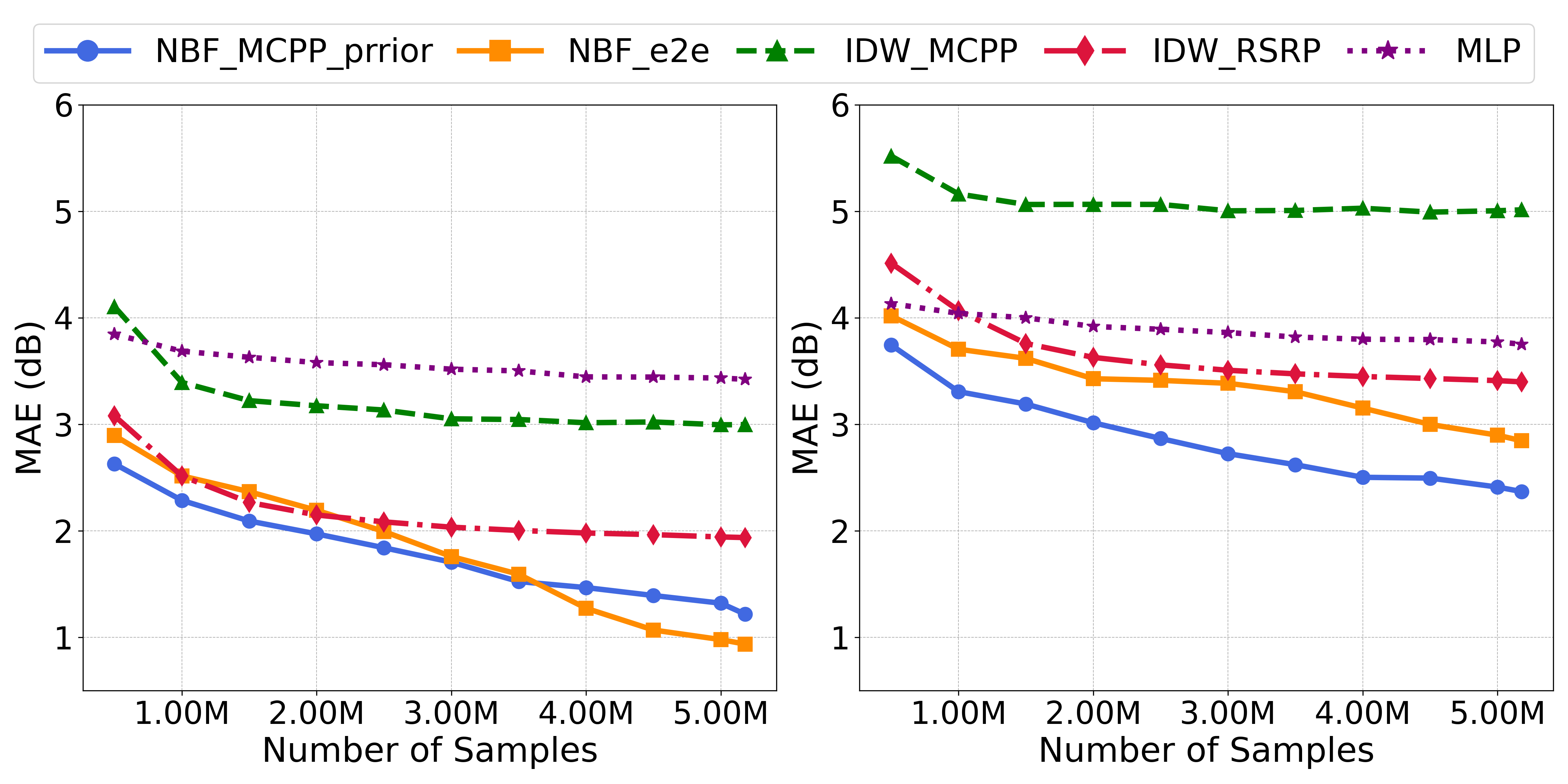}
 \caption{MAE comparison under ray-traced deterministic channel (left) and hybrid channel with add-on random MCPP components (right).}
 \label{fig:robustness_comparison}
\end{figure}

Finally, we evaluate the MAE of mean RSRP prediction under different number of available samples, as shown in Fig. \ref{fig:robustness_comparison}, under ray-traced deterministic channel (left) and hybrid channel with add-on random MCPP components (right).
In the deterministic scenario, both IDW methods perform well (better than MLP), with their accuracy improving as the number of samples grows. However, their performance degradates significantly in hybrid channels due to random channel factors, especially for the MCPP-based IDW method which relies on accurate MCPP priors.
In comparison, our NBF methods provides the overall best performance under both settings.
In particular, the MCPP-prior guided NBF performs consistently the best even under random perturbation of MCPP, which suggests that our PaC strategy effectively learns the physical groundings underneath the RSRP measurements and thus provides accurate/robust beam RSRP predictions under diverse practical settings.

\section{Conclusions}
We proposed the NBF, a hybrid neural-physical framework for spatial beam RSRP prediction, based on MCPP as an essential intermediary to amalgamate propagation channels and antenna/beam configurations. The derived closed-form RSRP statistics were validated through Monte Carlo simulations under ray-traced multipath environments. Compared with conventional table-based CKMs with fixed interpolation rules (e.g., IDW), and pure blackbox models such as MLP, our proposed NBF achieved superior prediction accuracy, better generalization, and higher storage efficiency. The Pretrain-and-Calibrate strategy further improved convergence and adaptability. These results highlight NBF as an accurate, interpretable, and efficient solution for beam management in dense wireless networks.
Future work may extend to wideband channels and hybrid beamforming schemes.

\appendices

\section{Proof of Proposition 1}\label{AppendixCSCG}


Based on \eqref{RSRPdef}, the normalized beam RSRP is given by
\begin{align}
    &P_\text{r}  =  |\textstyle\sum\nolimits_{l\in\mathcal{L}}\boldsymbol{g}_{l}^T\,\mathbf{w}_\text{tx}|^2 =|\textstyle\sum\nolimits_{l\in\mathcal{L}}A_l e^{j\Phi_l}|^2\notag \\
    & = \textstyle\sum\nolimits_{l\in\mathcal{L}} \textstyle\sum\nolimits_{m\in\mathcal{L}}  A_l A_m^* e^{j(\Phi_l - \Phi_m)} \notag\\
     &\stackrel{(a)}{=}   \textstyle\sum\nolimits_{l\in\mathcal{L}} \textstyle\sum\nolimits_{m\in\mathcal{L}}  \text{Re} \left\{ A_l A_m^* e^{j(\Phi_l - \Phi_m)}\right\} \notag \\
    &=   \textstyle\sum\nolimits_{l\in\mathcal{L}} \textstyle\sum\nolimits_{m\in\mathcal{L}}  P_{l,m} \cos{\Psi_{l,m}},
\end{align}
where $A_l \triangleq \sqrt{G(\Pi_{\text{tx},l})}\,\alpha_l\,e^{-j2\pi f_\text{c} \tau_l} \Delta_l
$, $\Delta_l \triangleq \boldsymbol{a}_\text{tx}^T(\Pi_{\text{tx},l})\,\mathbf{w}_\text{tx}$, $P_{l,m}\triangleq \sqrt{G(\Pi_{\text{tx},l}) G(\Pi_{\text{tx},m})}\,\alpha_l \alpha_m \left| \Delta_l \right| \left| \Delta_m \right|$, $\Psi_{l,m}\triangleq \Phi_l -\Phi_m + \angle\Delta_l - \angle\Delta_m- 2\pi f_\text{c} (\tau_{l} - \tau_{m})$, and the equation $(a)$ is due to $\textstyle\sum\nolimits_{l\in\mathcal{L}} \textstyle\sum\nolimits_{m\in\mathcal{L}} A_l A_m^* = \textstyle\sum\nolimits_{l\in\mathcal{L}} \textstyle\sum\nolimits_{m\in\mathcal{L}} \text{Re}(A_l A_m^*)$ in general.
Note that the random phases $\Phi_l,l\in\mathcal{L}$ appear in the term $\Psi_{l,m}$.
For the case with $l=m$, we have $\Psi_{l,m}=0$ and hence $\mathbb{E}\{\cos{\Psi_{l,m}}\}=1$.
For the case with $l\neq m$, the phase $\Psi_{l,m}$ follows the uniform random distribution in $[0,2\pi)$ due to the uniformly random phases $\Phi_l$ and $\Phi_m$, and hence $\mathbb{E}\{\cos{\Psi_{l,m}}\}=0$. Therefore, the mean RSRP is given by
\begin{align}
    &\mathbb{E}\{P_\text{r}\}  =\textstyle\sum\nolimits_{l\in\mathcal{L}}  G(\Pi_{\text{tx},l})\alpha_l^2 \left| \Delta_l \right|^2\\
    &=\textstyle\sum\nolimits_{l\in\mathcal{L}}  G(\Pi_{\text{tx},l})\alpha_l^2 \left| \boldsymbol{a}_\text{tx}^T(\Pi_{\text{tx},l})\,\mathbf{w}_\text{tx} \right|^2\notag\\
    &=\textstyle\sum\nolimits_{l\in\mathcal{L}} G(\Pi_{\text{tx},l}) \alpha_l^2 \left| \boldsymbol{a}_\text{tx,y}^T(\Pi_{\text{tx},l})\,\mathbf{w}_\text{tx,y} \right|^2 \left| \boldsymbol{a}_\text{tx,z}^T(\Pi_{\text{tx},l})\,\mathbf{w}_\text{tx,z} \right|^2/N_\text{tx}.\notag
\end{align}%
Based on \eqref{ay} and \eqref{wy}, the $y\text{-}$component of the array factor is
\begin{equation}
    \boldsymbol{a}_{\text{tx,y}}^T(\Pi_{\text{tx},l})\,\boldsymbol{w}_{\text{tx,y}}(\xi_{\text{tx,y}})
= S_{N_{\text{tx,y}}}(\zeta_{\text{tx,y},l} +\xi_{\text{tx,y}}),
\end{equation}
where $\zeta_{\text{tx,y},l}\triangleq \frac{2\pi f_\text{c} d_{\text{y}}}{c} \mathbf{h}^T \boldsymbol{u}_\text{tx}(\Pi_{\text{tx},l})$ and $S_N(\cdot)$ is defined as
\begin{equation}\label{SN}
S_N(\psi)
\triangleq \textstyle\sum_{n=0}^{N-1} e^{j n \psi}
= \frac{\sin(N\psi/2)}{\sin(\psi/2)}
e^{j (N-1)\tfrac{\psi}{2}}.
\end{equation}
Similarly, the $z\text{-}$component of the array factor is given by
\begin{equation}
    \boldsymbol{a}_{\text{tx,z}}^T(\Pi_{\text{tx},l})\,\boldsymbol{w}_{\text{tx,z}}(\xi_{\text{tx,z}})
= S_{N_{\text{tx,z}}}(\zeta_{\text{tx,z},l} +\xi_{\text{tx,z}}),
\end{equation}
where $\zeta_{\text{tx,z},l}\triangleq \frac{2\pi f_\text{c} d_{\text{z}}}{c} \mathbf{v}^T \boldsymbol{u}_\text{tx}(\Pi_{\text{tx},l})$.
Therefore, we have
\begin{equation}
    \mathbb{E}\{P_\text{r}\}=\textstyle\sum\nolimits_{l\in\mathcal{L}} G(\Pi_{\text{tx},l}) p_l |\Delta_l|^2=\textstyle\sum\nolimits_{l\in\mathcal{L}} \gamma_l,
\end{equation}%
where $\gamma_l\triangleq G(\Pi_{\text{tx},l}) p_l |\Delta_l|^2$ denotes the average power gain, and $|\Delta_l|^2= \frac{1}{N_\text{tx}} \big| S_{N_{\text{tx,y}}}(\zeta_{\text{tx,y},l} +\xi_{\text{tx,y}}) \big|^2 \big| S_{N_{\text{tx,z}}}(\zeta_{\text{tx,z},l} +\xi_{\text{tx,z}})\big|^2$ denotes the power gain of the array factor $\Delta_l$, for path $l\in\mathcal{L}$.

Similarly, the variance of $P_\text{r}$ is given by
\begin{align}
    &\text{Var}\{P_\text{r}\}  =\mathbb{E}\big\{(P_\text{r}-\mathbb{E}\{P_\text{r}\})^2\big\}\notag\\
    &=\mathbb{E}\big\{ \big(\textstyle\sum\nolimits_{l\in\mathcal{L}} \textstyle\sum\nolimits_{m\in\mathcal{L}\setminus\{l\}}  P_{l,m} \cos{\Psi_{l,m}}\big)^2\big\}\notag\\
    &\stackrel{(a)}{=}\textstyle\sum\nolimits_{l\in\mathcal{L}} \textstyle\sum\nolimits_{m\in\mathcal{L}\setminus\{l\}}  2 P_{l,m}^2 \mathbb{E}\{ (\cos{\Psi_{l,m}})^2\}\notag\\
    &\stackrel{(b)}{=}\textstyle\sum\nolimits_{l\in\mathcal{L}} \textstyle\sum\nolimits_{m\in\mathcal{L}\setminus\{l\}}  \gamma_l\gamma_m = \left(\textstyle\sum\nolimits_{l\in\mathcal{L}} \gamma_l\right)^2 - \textstyle\sum\nolimits_{l\in\mathcal{L}} \gamma_l^2,
\end{align}%
where the equation $(a)$ is due to $P_{l,m}\cos{\Psi_{l,m}}=P_{m,l}\cos{\Psi_{m,l}}$ and $\mathbb{E}\{ \cos{\Psi_{l,m}}\}=0$ for $l\neq m$, and $(b)$ is due to $P_{l,m}^2=\gamma_l\gamma_m$ and $\mathbb{E}\{ (\cos{\Psi_{l,m}})^2\}=1/2$. \textbf{Proposition 1} thus follows. $\blacksquare$

\bibliography{IEEEabrv,BibDIRP,beam}

\newpage

\end{document}